\documentclass[]{spie}  
\usepackage[]{graphicx,journals}

\title{An aperture masking mode for the MICADO
instrument} 

\author{S. Lacour\supit{a}, P. Baudoz\supit{a}, E. Gendron\supit{a}, A. Boccaletti\supit{a}, R. Galicher\supit{a}, Y. Cl\'enet\supit{a}, D. Gratadour\supit{a},  T. Buey\supit{a}, G. Rousset\supit{a}, M. Hartl\supit{b} and R. Davies\supit{b}
\skiplinehalf
\supit{a} LESIA/Observatoire de Paris, 5 place Jules Janssen, 92195 Meudon,
France \\
\supit{b} Max-Planck Institut fur extraterrestrische Physik (MPE), 85748, Garching, Germany \\
}

\authorinfo{For further information, please send correspondence to: sylvestre.lacour@obspm.fr}

\begin{document} 
  \maketitle 
\begin{abstract}
MICADO is a near-IR camera for the European
ELT, featuring an extended field (75" diameter) for imaging, and also
spectrographic and high contrast imaging capabilities. It has been chosen
by ESO as one of the two first-light instruments. Although it is ultimately
aimed at being fed by the MCAO module called MAORY, MICADO will come
with an internal SCAO system that will be complementary to it and will
deliver a high performance on axis correction, suitable for coronagraphic
and pupil masking applications. The basis of the pupil masking approach
is to ensure the stability of the optical transfer function, even in the case
of residual errors after AO correction (due to non common path errors and
quasi-static aberrations). Preliminary designs of pupil masks are presented.
Trade-offs and technical choices, especially regarding redundancy and
pupil tracking, are explained.
\end{abstract}



\section{INTRODUCTION}
\label{sec:intro}  

   \begin{figure}
   \begin{center}
   \begin{tabular}{c}
   \includegraphics[height=12cm]{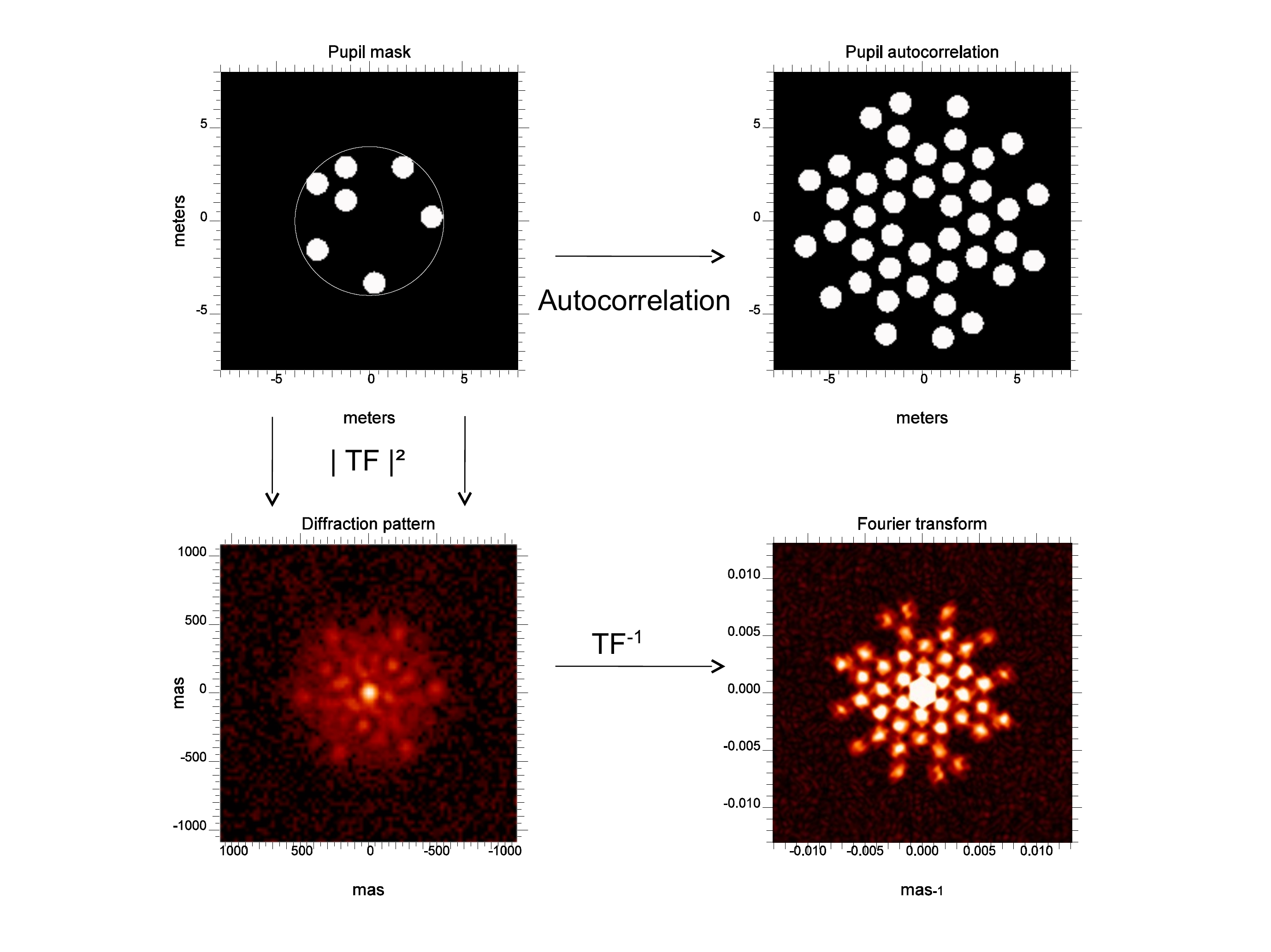}
   \end{tabular}
   \end{center}
   \caption[example] 
   { \label{fig:principe} 
Basic principles about aperture masking and Fourier optics. According to Fraunhofer diffraction, the electric field in the focal plane can be calculated by taking the Fourier transform of the electric field in the pupil plane. Moreover, the image in the focal plane is the intensity of this electric field. Therefore, the image is the power of the Fourier transform of the pupil mask. This is represented in this figure by the lower left image which was obtained on the NACO instrument at the VLT. The upper left image is a representation of the 7-hole mask used on that instrument. The lower right image is the Fourier transform of the image: it corresponds to the spatial frequencies present in the image. The spatial frequencies allowed by the mask are the baselines formed between the holes. These spatial frequencies can be calculated by taking the autocorrelation of the pupil mask (upper right panel). }
   \end{figure} 

The optical transfer function (OTF) of a telescope is the autocorrelation of the electric field in the pupil of the telescope (see figure~\ref{fig:principe}). The principle of  non-redundant masking (NRM) is to insert in the pupil an opaque mask with a number of holes.
The idea is that each pair of holes will contribute to a unique spatial frequency. The goal of NRM is to avoid the combinations of multiple spatial frequencies which -- adding incoherently -- may degrade transfer function.

Deep down, the power of NRM is to be able to distinguish the effect of optical aberrations from astronomical information. This can be done for example by computing the closure phase. The closure phase is unaffected by phase errors in the pupils, but modified by the asymmetry of the astronomical target\cite{2011Msngr.146...18L}. However, this concept of NRM can be extended to partially redondant masking. This is the principle of Kernel phase\cite{2010ApJ...724..464M}. The idea is to introduce only {\em some} degree of non-redundancy\cite{2013MNRAS.433.1718I}. 

Aperture masking has proved its complementarity with traditional AO systems. Even on modern instruments, with low optical aberrations, non-redundant aperture masks are present (NIRISS on JWST\cite{2012SPIE.8442E..2SS}, SPHERE\cite{2008SPIE.7014E..41B} on the VLT, GPI\cite{2008SPIE.7015E..31M} at GEMINI observatory). The reason for that is that aperture masking offers at no cost resolution capabilities which go below the traditional diffraction limit of $1.22\,\lambda/D$. More exactly, it gives constant dynamic range down to $\lambda /2 B$, where $B$ is maximum baseline length in the pupil  ($B \leq D$)\cite{2011A&A...532A..72L}. Observations on traditional AO system (like NACO on the VLT)\cite{2008ApJ...679..762K,2011ApJ...731....8K,2012ApJ...744..120E}, showed that NRM routinely gives dynamic range of the order of 7 magnitudes in the K band.

In the case of the E-ELT, NRM -- or partially non-redundant masking -- have specific advantages due to such huge segmented aperture. First because the cophasing of the 798 mirrors may not be perfect, causing phase jumps difficult to deconvolve by classical imaging. Second, because vibrations of the segments can be difficult to correct with AO systems. Last, because segments can be missing. For these additional reasons which are specific to the E-ELT, we  designed masks for MICADO which will make an intended compromise between redundancy and flux throughput.

\section{MICADO AND PRELIMINARY REQUIREMENTS}
\label{sec:base}

\begin{figure}
   \begin{center}
   \begin{tabular}{c}
   \includegraphics[height=6cm]{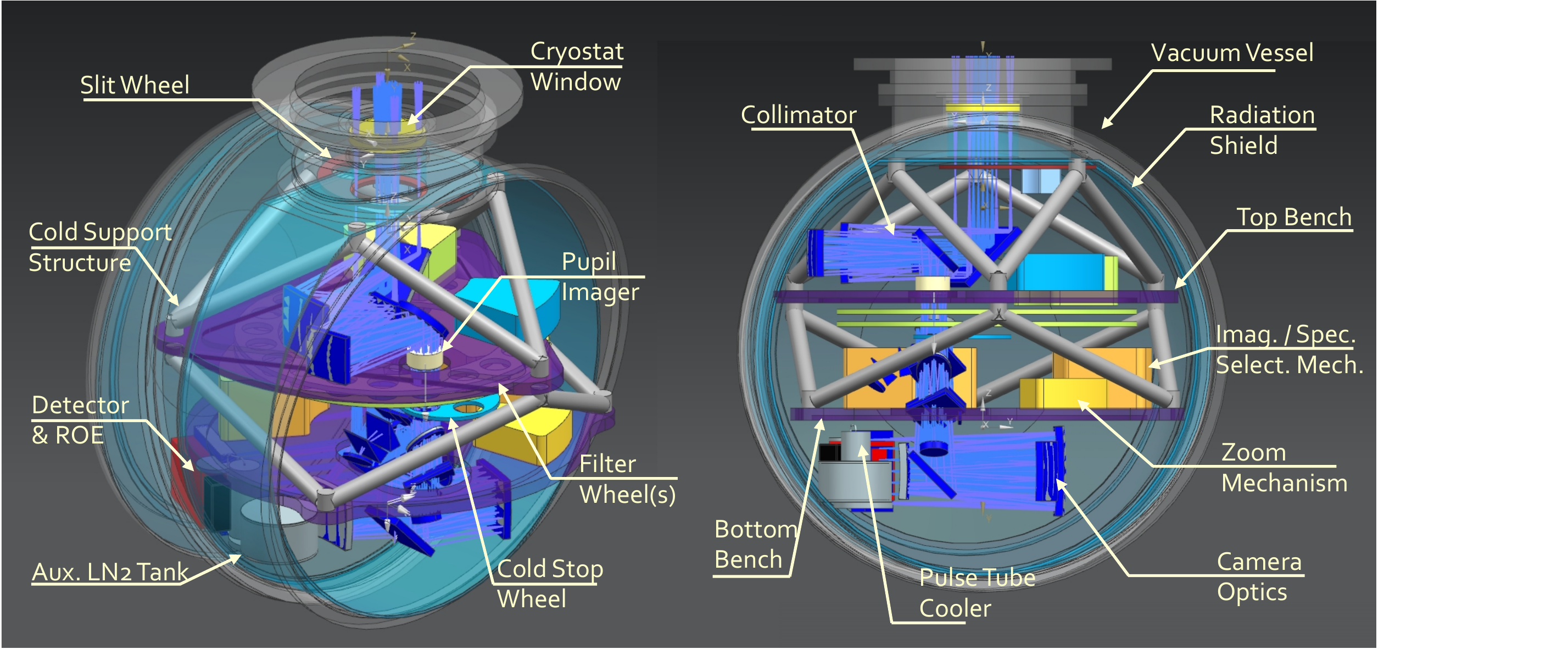}
   \end{tabular}
   \end{center}
   \caption[example] 
   { \label{fig:micado} 
Preliminary optical layout of the MICADO instrument}
   \end{figure}

MICADO\cite{2010SPIE.7735E..77D} is the first-light near-infrared camera (0.8-2.5\,$\mu$m) on the E-ELT. It has been designed to work at the diffraction limit over a $\approx1$' field of view and will come with a long slit spectroscopic mode, at a moderate spectral resolution (4000 to 8000, possibly 20000). Figure~\ref{fig:micado} presents a preliminary layout of the different sub-systems inside MICADO. A cold stop wheel is present inside the cryostat, where the mask will be inserted.

The requirements of aperture masking can be divided into two groups:  the pupil requirements and on the focal plane requirements.
\begin{itemize}
\item on the pupil. The main requirement is to avoid the spider arms, as well as gaps between segments. This means that this mode will require pupil tracking (the instrument rotator follows the pupil of the telecope, instead of the rotation of the sky). The M1 segments are about 1.45\,m corner to corner, i.e. about 1.25\,m flat edge to flat edge. For 84\,cm diameter holes, it means that the hole can move by a margin of 41\,cm without touching the edge of the segment. Hence it gives a requirement on the stability of the pupil: the pupil of the telescope (or more exactly, the M1 mirror), must be positioned within 1\% with respect to the pupil wheel inside the cryostat. In terms of pupil rotation tracking, this also means a maximum angular error of atan$(41/3900)=0.6$\,degree.
\item on the size and sampling of the focal plane. The pixel sampling of the MICADO instrument is 3\,mas. This correspond to $0.5 \lambda/D$ at 1\,$\mu$m, the angular resolution of the aperture mask, and Shannon's frequency. It means that for a wavelength shorter than 1\,$\mu$m, the mask will not be available, or at least not without having to discard the highest spatial frequencies.
In terms of required size of detector, with $d=0.84$\,m holes, the diffraction pattern will cover an area equal to the Airy pattern of a single hole. At the longest wavelength, i.e. K band, this corresponds to $2\times 1.22\,\lambda/d\approx1\,$arcseconds, which a detector area of $512\times512$ pixels would be sufficient to cover.
\item on the readout speed. With coherence times of the order of the milliseconds, it is useless to try to correct the atmosphere by using NRM techniques. But anyway, the objective of correcting the atmosphere is reserved to the adaptive optics system. The goal of the aperture mask will be to correct the remaining non-common path errors and/or quasi-static errors. To do so, the faster the camera will acquire the images, the more resilient the system will be. Our objective is to be able to catch ELT vibrations, which may be one of the principal source of concern for the segmented aperture. The vibrations eigenfrequencies have been calculated for the ELT, with the 6 principal modes ranging from 2 to 5\,Hz. Hence, to sample these vibrations, we would need a frame rate above 10\,Hz, i.e., integration times smaller than 100\,ms.
\end{itemize}

\section{FULLY NON-REDUNDANT MASK}
\label{sc:nrm}

   \begin{figure}
   \begin{center}
   \begin{tabular}{c}
   \includegraphics[width=5cm]{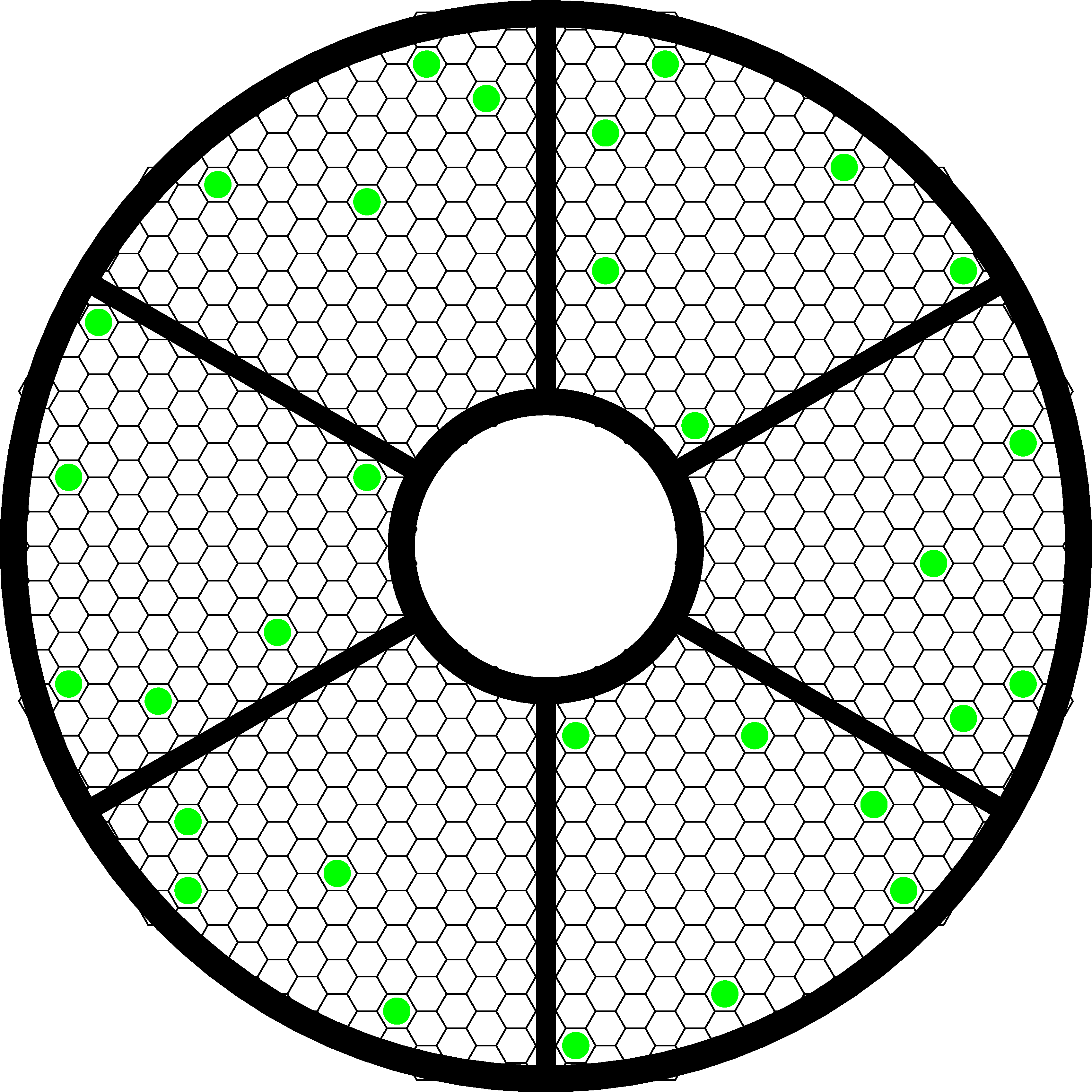}
   \includegraphics[width=5.5cm]{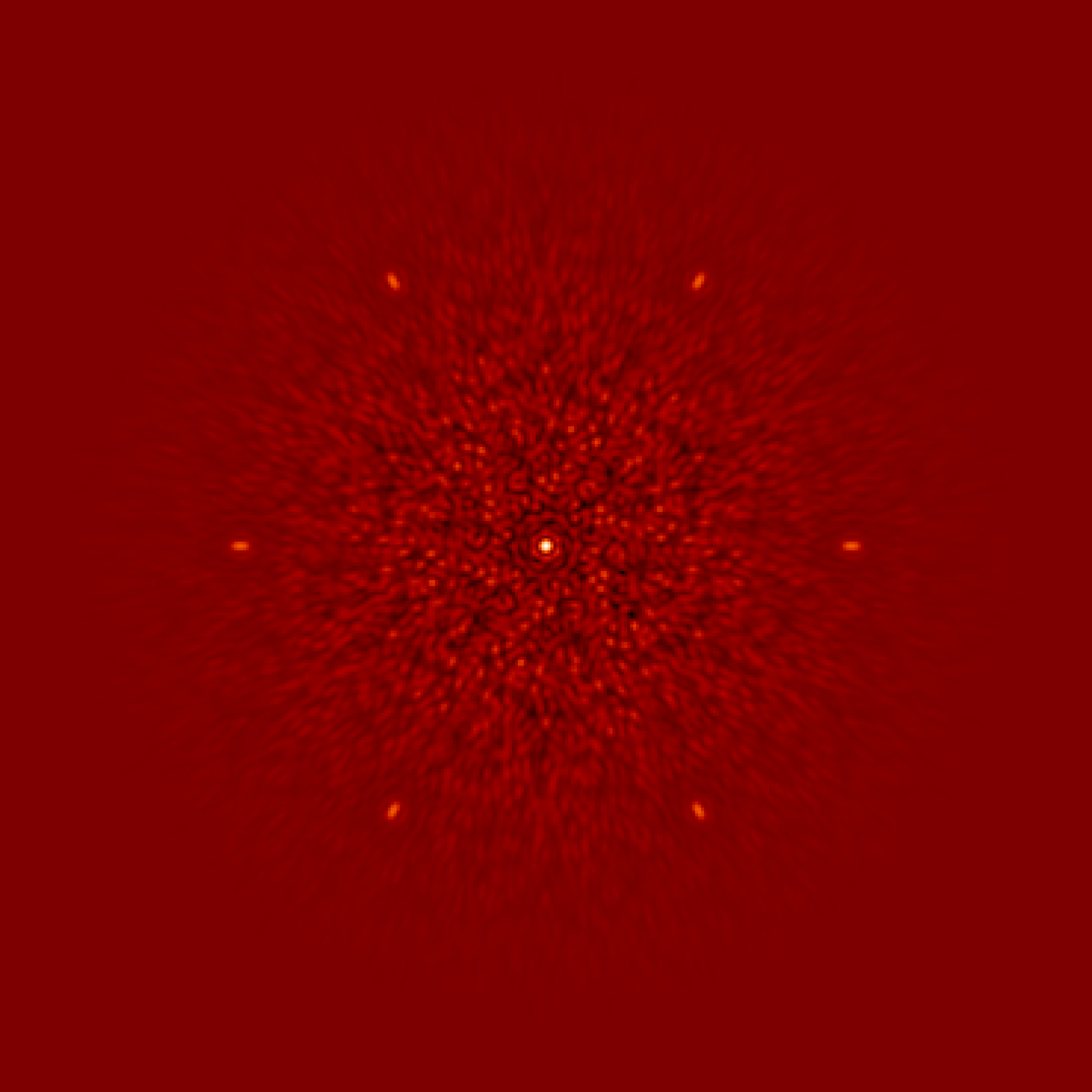}\\
   \includegraphics[width=5cm]{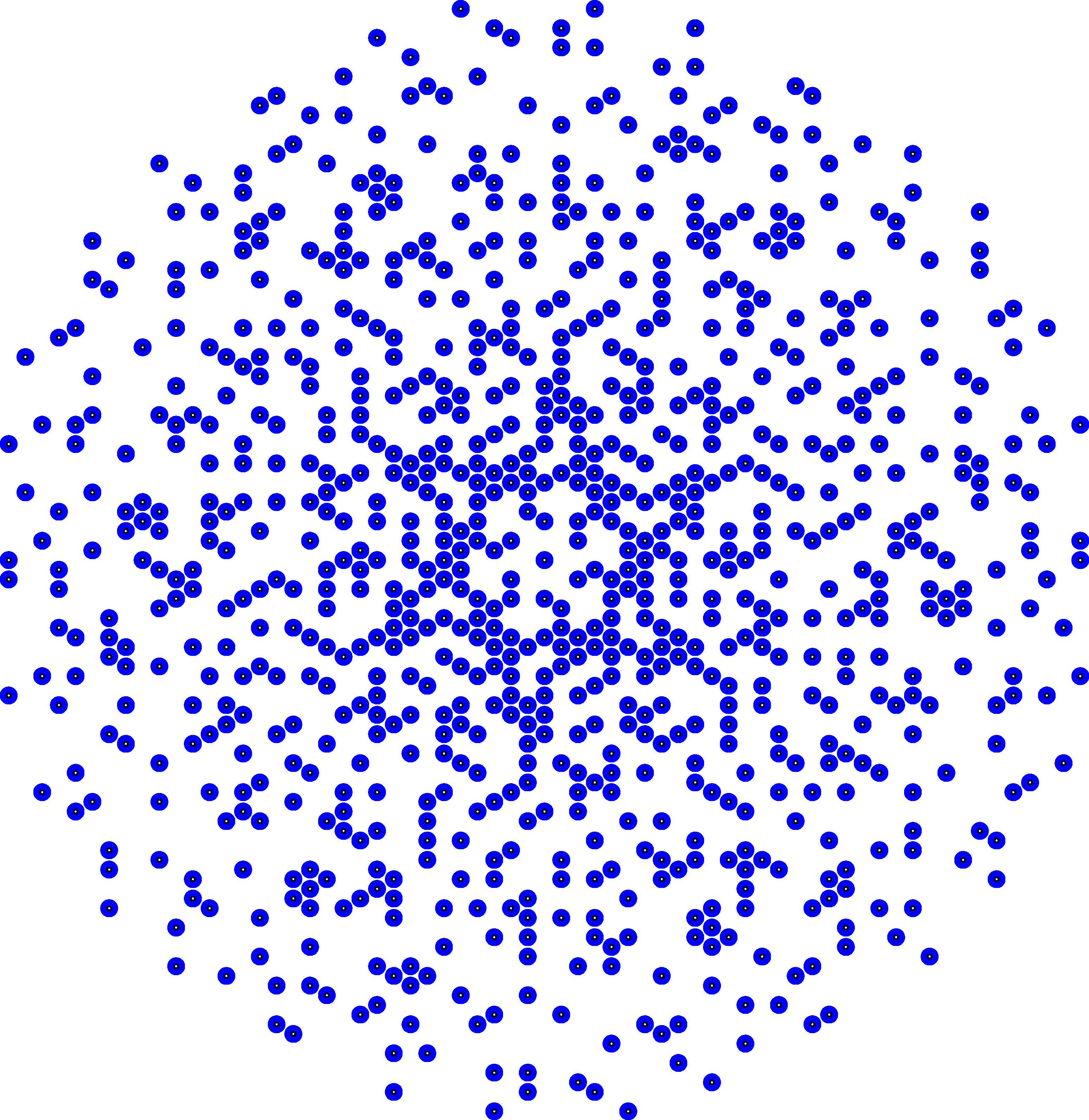}
   \includegraphics[width=5.5cm]{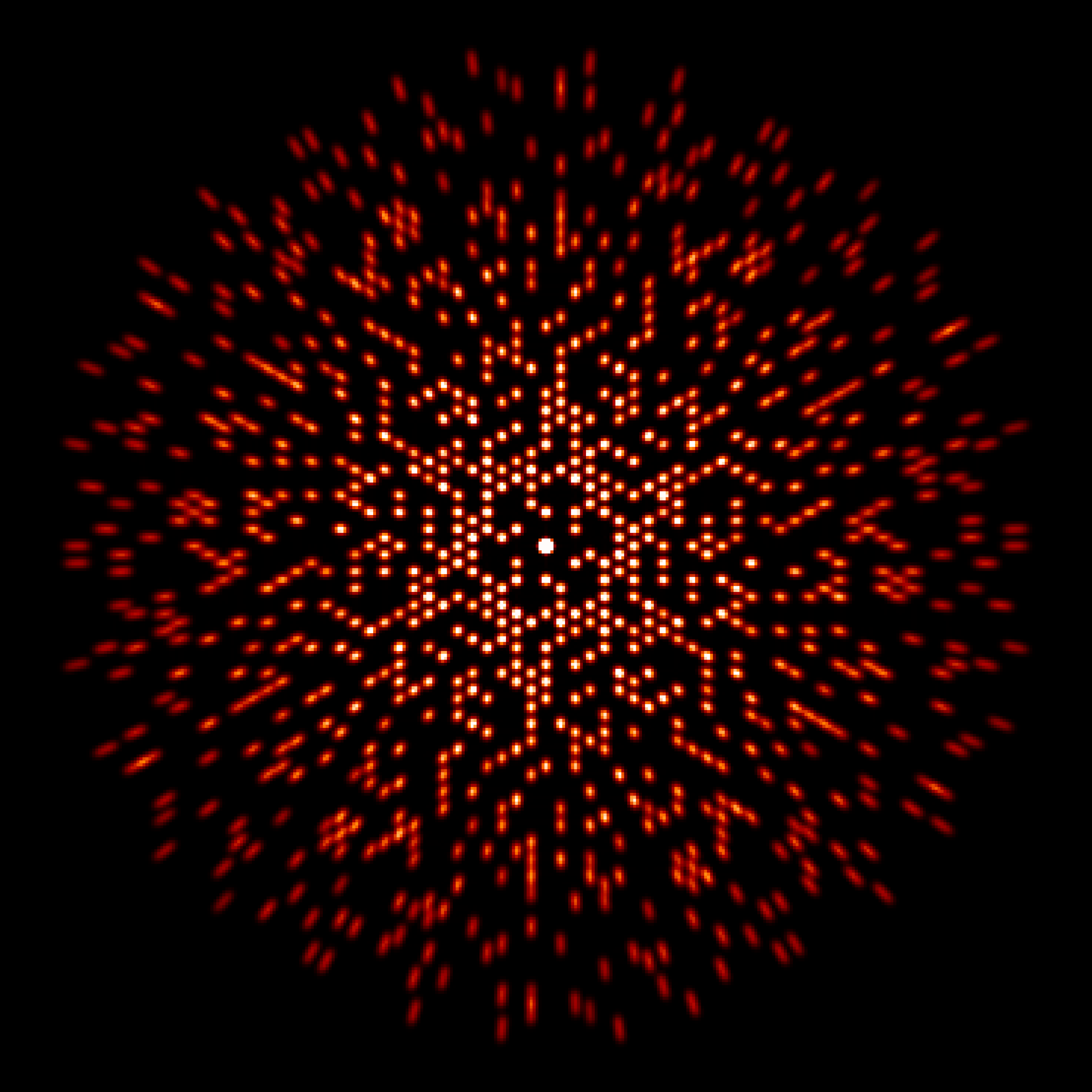}
   \end{tabular}
   \end{center}
   \caption[example] 
   { \label{fig:NRM} 
The fully non-redundant mask. Top left panel: in green are represented the holes of the aperture mask which will correspond to a designated segment. Upper right panel: simulated PSF (K band, 750\,mas wide). Lower left panel: the $u$-$v$ coverage, with the zero spatial frequency in red. Lower right panel: Fourier transform of the simulated PSF for broad band (K band) observations.}
   \end{figure} 

There are no perfect non-redundant mask. Many possibilities exists. The design of a NRM mask must be a compromise between several factors:
\begin{enumerate}
\item flux throughput, a balance between the number of holes $N$, and the size of the holes $d$. Evidently, the throughput is proportional to $N\times d^2$.
\item $u$-$v$ coverage. The more holes, the more spatial frequencies. With a NRM mask, the spatial frequencies are equal to $N\times(N-1)/2$. However, a good repartition of the spatial frequencies is also needed to sample homogeneously the $u$-$v$ plane, especially including the lowest and highest frequencies.
\item frequency overlapping due to spectral smearing. The spatial frequency is proportional to $B/ \lambda$. For the longest baselines, the spectral bandpass can significantly extend radially the spatial frequencies. Therefore, we will try to avoid having the longest baselines matching too closely each other (at least radially).
\end{enumerate}

The diameter of the holes is determined by the minimum distance between spatial frequencies. The need is to minimize the overlapping between the frequencies allowed within a hole ($d/\lambda$) and between two pairs of holes ($\min (B_i-B_j) / \lambda$). This distance between the spatial frequencies $B_i-B_j$ correspond to the spacing of an hexagonal grid in the frequency domain. In the case of the ELT, it corresponds to the hexagonal grid on which are placed the segments of the primary mirror. The spacing of the grid is therefore 1.25\,m in unit of wavelength. To fully avoid redundancy, the diameter of the holes should be half that distance, i.e, below 63\,cm. However, there are some tolerance to this ratio, to increase the throughput. For NACO, the minimum distance between two holes is 1.8\,m for diameters of 1.2\,m\cite{2010SPIE.7735E..56T}, hence a ratio 1.5 instead of 2. Using the same ratio of 1.5, we derive holes of size 84\,cm. We can comment of the arbitrary nature of this value 1.5. For example for the SPHERE instrument, this value has been scaled up to 1.6, with holes of diameter 1.1\,m.

Once the size of the holes is decided, as well as the hexagonal grid (deduced from the position of the segments), the repartition of the holes on the ELT pupil has to been done to i) conserve the non-redundancy, ii) maximize the number of holes, iii) avoid as much as possible the overlapping of spatial frequency due the spectral bandpass. The mask presented in Figure~\ref{fig:NRM} have been found according to these three criteria. The number of holes is 30, representing slightly more than 3.7\% of the segments.

\section{PARTIALLY NON-REDUDANT MASK}

   \begin{figure}
   \begin{center}
   \begin{tabular}{c}
   \includegraphics[width=5cm]{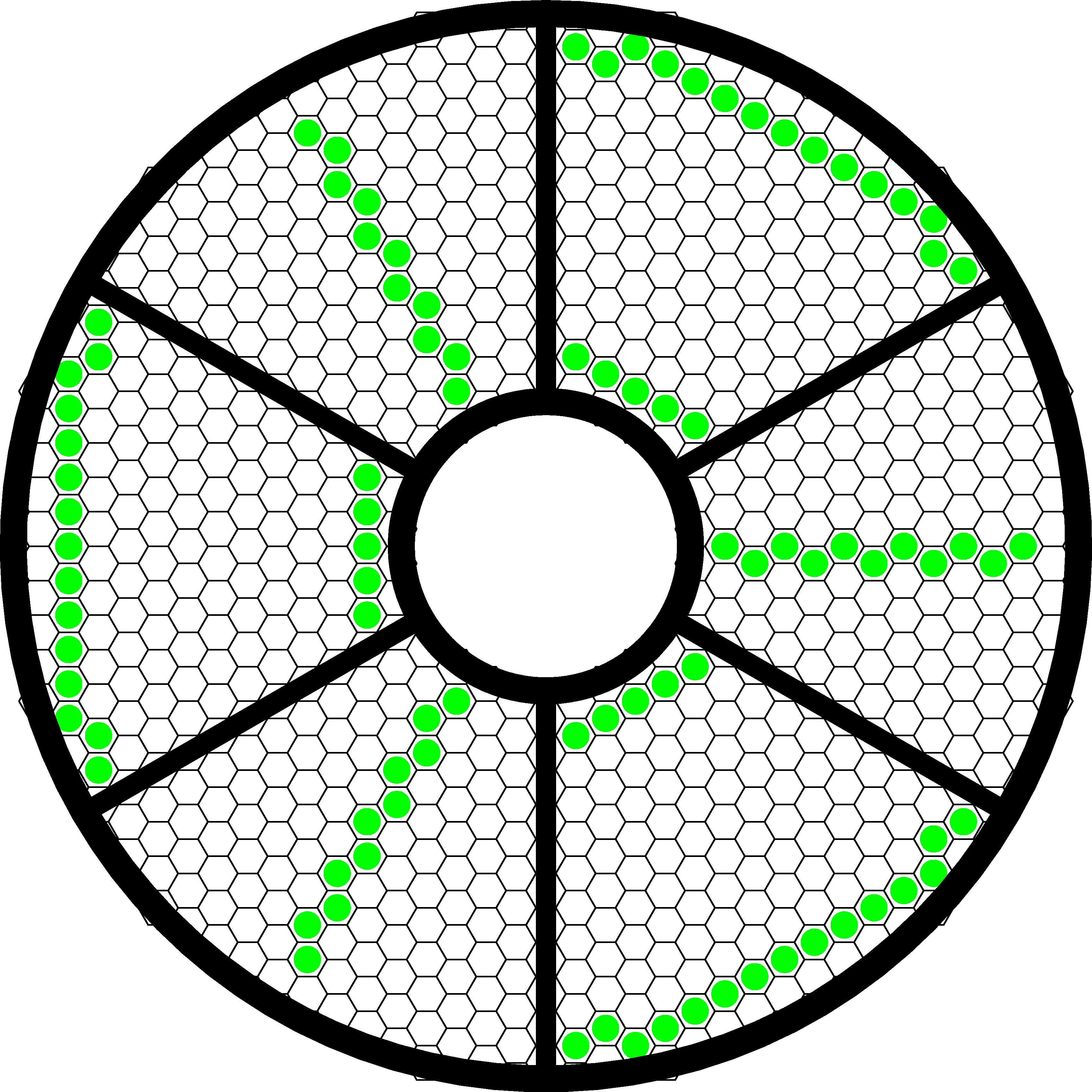}
   \includegraphics[width=5.5cm]{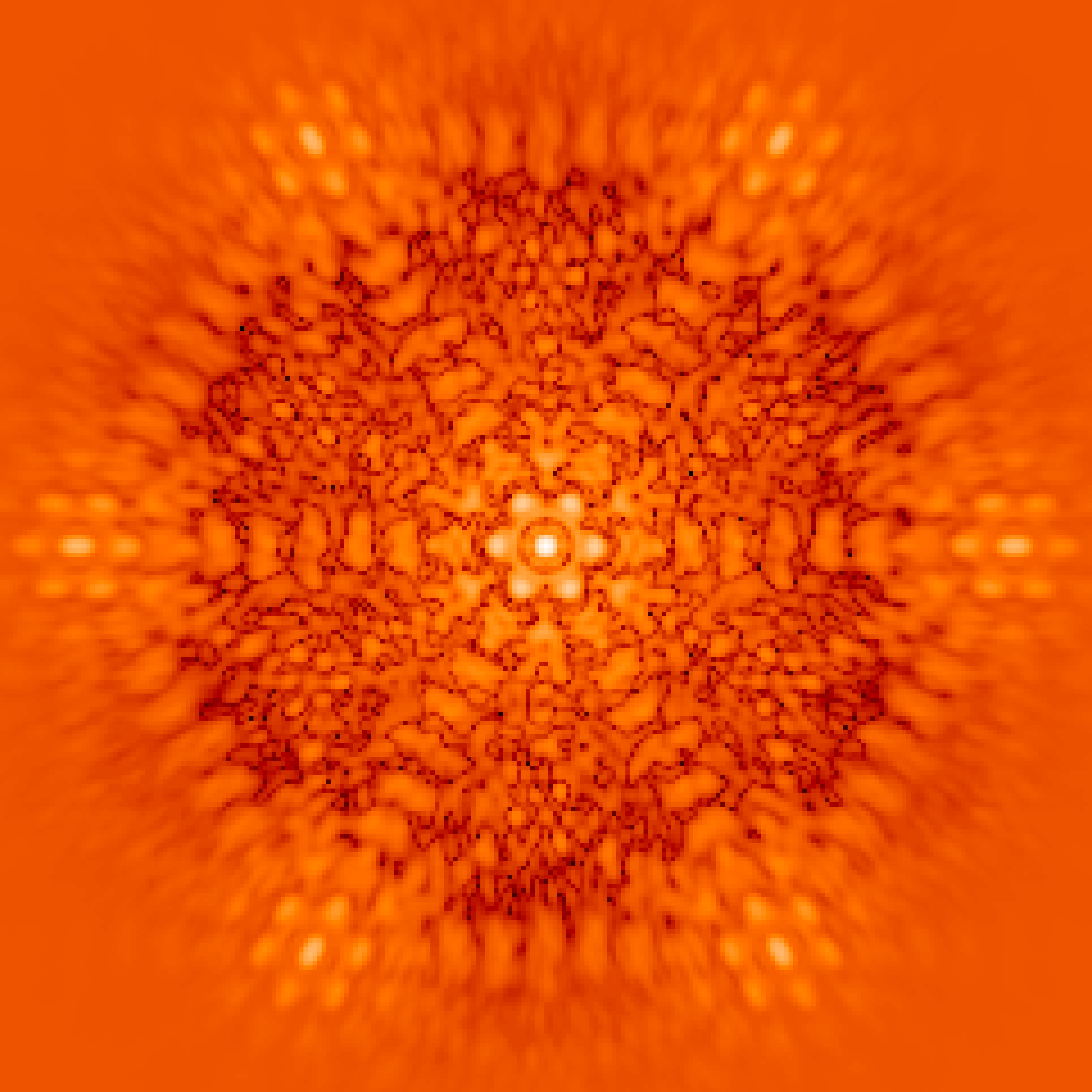}\\
   \includegraphics[width=5cm]{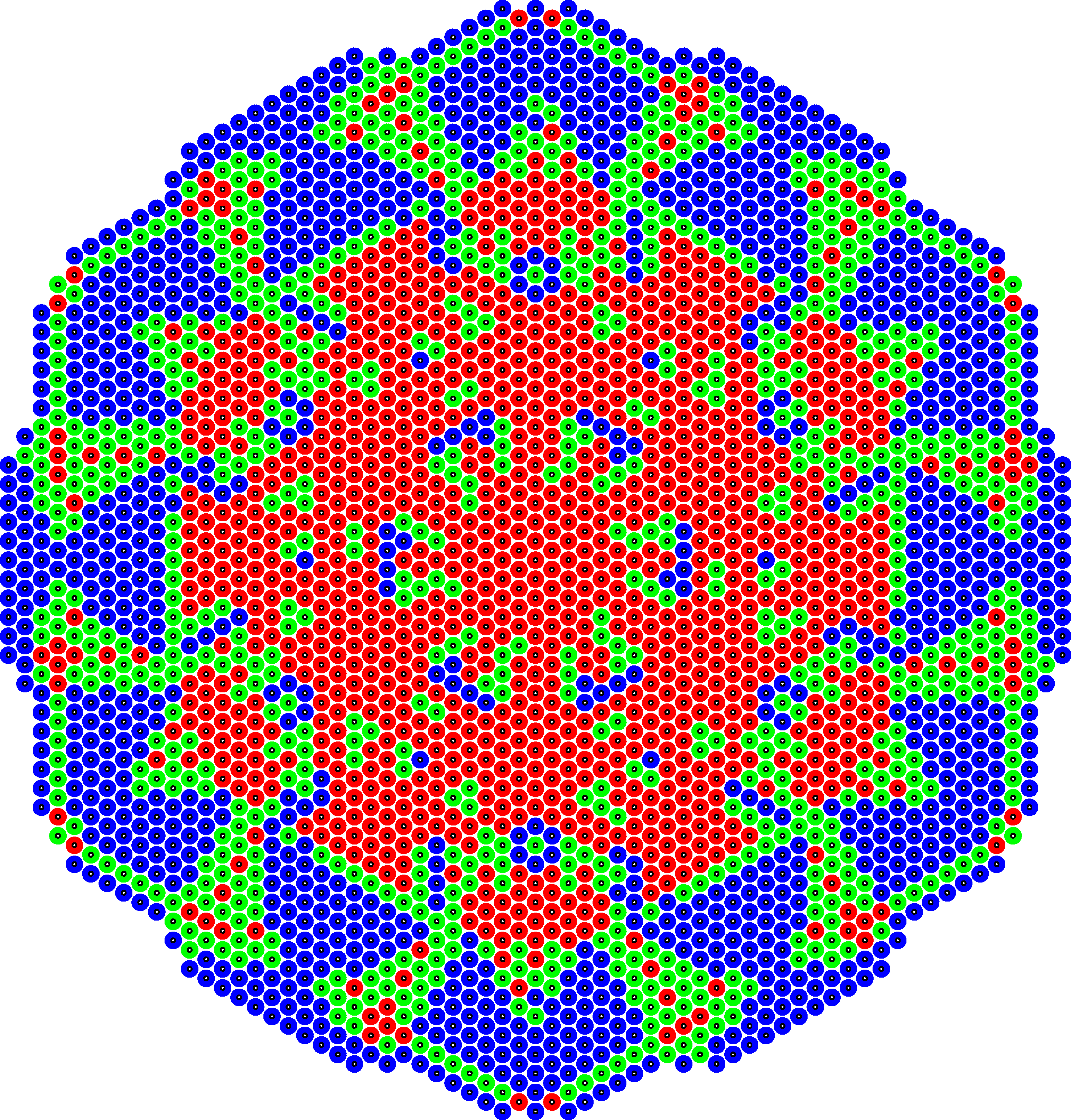}
   \includegraphics[width=5.5cm]{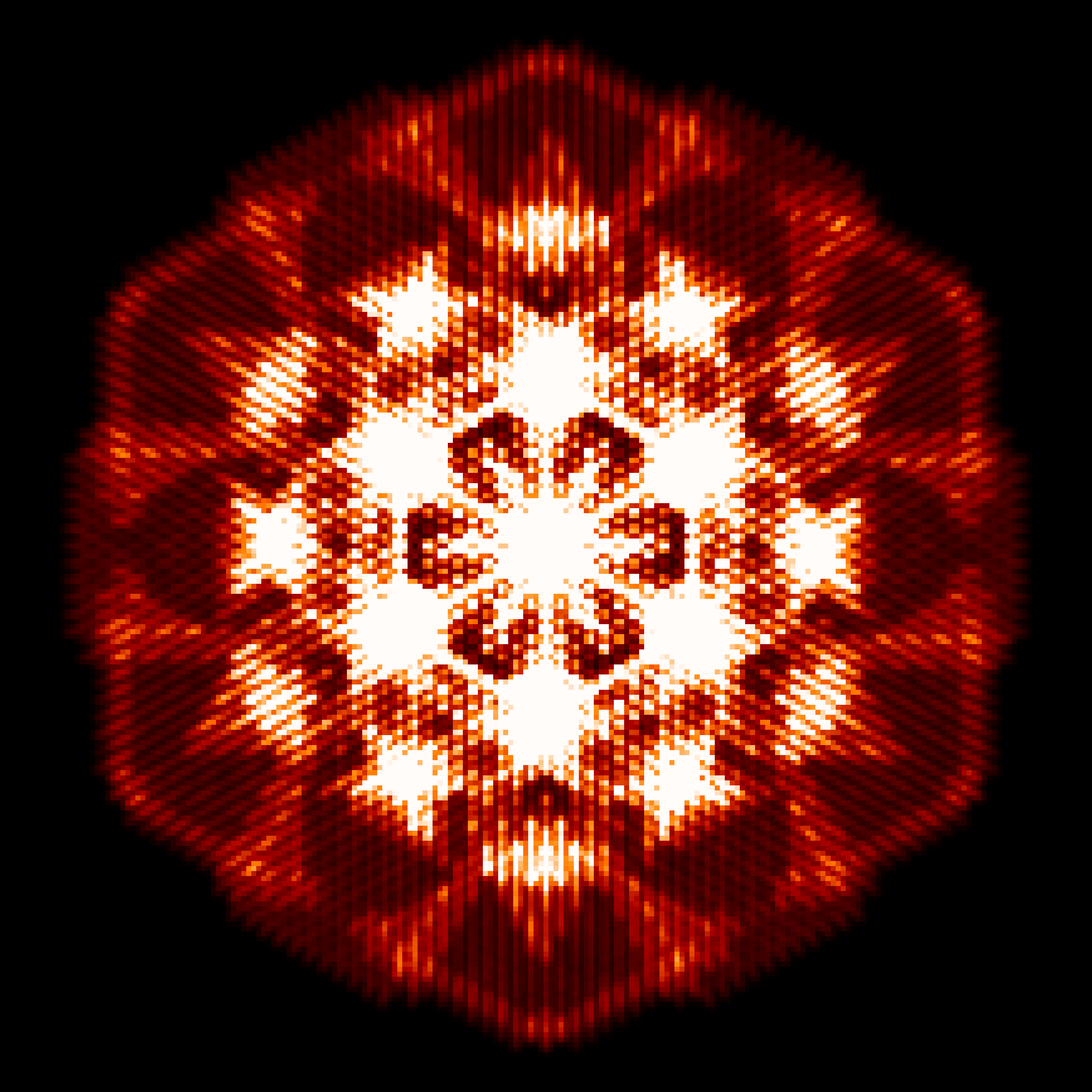}
   \end{tabular}
   \end{center}
   \caption[example] 
   { \label{fig:PNRM} 
Partially non-redundant mask. Represented are the same panel as in Figure~\ref{fig:NRM}. The $u$-$v$ coverage in the lower left panel is represented with colors. The blue markers correspond to spatial frequencies which correspond to a unique pair of holes. The green points marks  spatial frequencies which are present twice in the pupil. And in red, frequencies which are caused by more than two pairs.}
   \end{figure} 

The choice of an NRM mask is the fruit of compromises stated in section~\ref{sc:nrm}. By relaxing the constraint on the redundancy, we can further gain in terms of throughput and $u$-$v$ coverage.  The theory of phase deconvolution through kernels  has been mathematically established by Martinache et al.\cite{2010ApJ...724..464M}. The kernel phase formalism is working well in the case of small perturbations. It assumes that the phase measured at a given spatial frequency is a linear combination of the phases in the pupil which contribute to that spatial frequency. Kernel phase is then based on the fact that it is possible to take the Kernel of the contribution of the pupil phase. The non-zero kernel phase is then caused by the asymmetry of the astronomical object. This technique is similar to the closure phase technique on fully non-redundant masks, but permits an extention to partially non-redundant masks. Such an idea is perfectly valid in the case of an ELT\cite{2013aoel.confE...6M}. 

Thus, the second mask for MICADO is partially non-redundant and is presented in Figure~\ref{fig:PNRM}. Compared to the other mask, it will have a higher througput: 93 segments will be used instead of 30 for the perfectly non-redundant mask. Over the 1462 spatial frequencies probed by the mask, 519 are uniques (i.e., are due to a unique pair of holes in the mask). They are represented in blue in the lower left panel of Figure~\ref{fig:PNRM}. The mask is therefore at $519/1461 \approx 35 \%$ non-redudant. 384 spatial frequencies are caused by two baselines, 201 by three baselines, and the remaining 357 by 4 to 28 baselines.

The main particularity of this mask is a complete coverage of the $u$-$v$ space, up to 38\,m. This total coverage will enable direct image reconstruction, without the need for a prior or an assumption on the observed object.

\section{LAST REMARK}

Aperture masking gives imaging capabilities at twice the angular resolution of a telescope. Combined with the high dynamic range already demonstrated on 8-meter class telescopes, it will offer uniques capabilities to the ELT. More specifically, it will most certainly deliver groundbreaking results on young stellar objects and evolved stars.

The implementation of aperture masking is rather simple and inexpensive. However, \textbf{the most stringent requirement of this mode will be pupil stabilization.} If the pupil moves by more than 1\%, the performances will be severely degraded. In extreme cases, the holes can be vignetted (example of severe vignetting are observed on GPI by Greenbaum et al.\cite{2014arXiv1407.2310G}). However, it has to be said that pupil stabilization is an obvious requirement for any instrument willing to do coronagraphy and/or high dymanic range imaging 

\acknowledgments     

 This work was supported the French National Agency for Research (ANR-13-JS05-0005-01). 


\bibliography{mica}   
\bibliographystyle{spiebib}   

\end{document}